\documentstyle[]{mn}

\newif\ifAMStwofonts

\ifoldfss
  \ifCUPmtlplainloaded \else
    \NewTextAlphabet{textbfit} {cmbxti10} {}
    \NewTextAlphabet{textbfss} {cmssbx10} {}
    \NewMathAlphabet{mathbfit} {cmbxti10} {} 
    \NewMathAlphabet{mathbfss} {cmssbx10} {} 
  \fi
  \ifAMStwofonts
    \ifCUPmtlplainloaded \else
      \NewSymbolFont{upmath} {eurm10}
      \NewSymbolFont{AMSa} {msam10}
      \NewMathSymbol{\upi}     {0}{upmath}{19}
      \NewMathSymbol{\umu}     {0}{upmath}{16}
      \NewMathSymbol{\upartial}{0}{upmath}{40}
      \NewMathSymbol{\leqslant}{3}{AMSa}{36}
      \NewMathSymbol{\geqslant}{3}{AMSa}{3E}

      \let\leq=\leqslant 
      \let\geq=\geqslant 
    \fi
  \fi
\fi 

\ifnfssone
  \newmathalphabet{\mathit}
  \addtoversion{normal}{\mathit}{cmr}{m}{it}
  \addtoversion{bold}{\mathit}{cmr}{bx}{it}
  \newmathalphabet{\mathbfit} 
  \addtoversion{normal}{\mathbfit}{cmr}{bx}{it}
  \addtoversion{bold}{\mathbfit}{cmr}{bx}{it}
  \newmathalphabet{\mathbfss} 
  \addtoversion{normal}{\mathbfss}{cmss}{bx}{n}
  \addtoversion{bold}{\mathbfss}{cmss}{bx}{n}
  \ifAMStwofonts
    \ifCUPmtlplainloaded \else
      \UseAMStwoboldmath
      \makeatletter
      \new@mathgroup\upmath@group
      \define@mathgroup\mv@normal\upmath@group{eur}{m}{n}
      \define@mathgroup\mv@bold\upmath@group{eur}{b}{n}
      \edef\UPM{\hexnumber\upmath@group}
      \new@mathgroup\amsa@group
      \define@mathgroup\mv@normal\amsa@group{msa}{m}{n}
      \define@mathgroup\mv@bold\amsa@group{msa}{m}{n}
      \edef\AMSa{\hexnumber\amsa@group}
      \makeatother
      \mathchardef\upi="0\UPM19
      \mathchardef\umu="0\UPM16
      \mathchardef\upartial="0\UPM40
      \mathchardef\leqslant="3\AMSa36
      \mathchardef\geqslant="3\AMSa3E

      \let\leq=\leqslant 
      \let\geq=\geqslant 
    \fi
  \fi
\fi 

\ifnfsstwo
  \DeclareMathAlphabet{\mathbfit}{OT1}{cmr}{bx}{it}
  \SetMathAlphabet\mathbfit{bold}{OT1}{cmr}{bx}{it}
  \DeclareMathAlphabet{\mathbfss}{OT1}{cmss}{bx}{n}
  \SetMathAlphabet\mathbfss{bold}{OT1}{cmss}{bx}{n}
  \ifAMStwofonts
    \ifCUPmtlplainloaded \else
      \DeclareSymbolFont{UPM}{U}{eur}{m}{n}
      \SetSymbolFont{UPM}{bold}{U}{eur}{b}{n}
      \DeclareSymbolFont{AMSa}{U}{msa}{m}{n}
      \DeclareMathSymbol{\upi}{0}{UPM}{"19}
      \DeclareMathSymbol{\umu}{0}{UPM}{"16}
      \DeclareMathSymbol{\upartial}{0}{UPM}{"40}
      \DeclareMathSymbol{\leqslant}{3}{AMSa}{"36}
      \DeclareMathSymbol{\geqslant}{3}{AMSa}{"3E}

      \let\leq=\leqslant 
      \let\geq=\geqslant 
    \fi
  \fi
\fi 

\ifCUPmtlplainloaded \else
  \ifAMStwofonts \else 
    \def\upi{\pi}
    \def\umu{\mu}
    \def\upartial{\partial}
  \fi
\fi

\title{Shocks and dust survival in nearby active galaxies: implications for the
alignment effect}
\author[M. Villar-Mart\'\i n  et al.]
       {M. Villar-Mart\'\i n$^1$,  D. De Young$^2$, A. Alonso-Herrero$^{1,5}$, M. Allen$^3$, L. Binette$^4$\\
        $^1$ Dept. of Natural Sciences, University of Hertfordshire, College Lane,
Hatfield-Herts, AL10 9AB \\
	$^2$ National Optical Astronomy Observatory,  950 North Cherry Avenue, P.O. Box 26732, Tucson, AZ 85726,
USA \\
	$^3$ Space Telescope Science Institute, 3700 San Martin Drive, Baltimore, Maryland 21218, USA\\
	$^4$ Instituto de Astronom\'\i a, UNAM, Apartado Postal 70-264,  D.F. 04510,  MEXICO \\
	$^5$ Present address: Steward Observatory, The University of Arizona, Tucson, AZ 85721, USA
}
\date{}

\pagerange{\pageref{firstpage}--\pageref{lastpage}}
\pubyear{2001}

\begin{document}

\maketitle

\label{firstpage}

\begin{abstract}

	One of the most popular explanations for the  so-called alignment 
effect  in high redshift ($z>$0.7) radio galaxies is the scattering by dust
of the hidden quasar light. 
As shown by De Young (1998)  a problem with the dust scattering model is that the
short destruction time-scale for dust grains means that they will not
survive the passage of the radio jet. 

	 We  investigate  the survival of dust in
the extended ionised gas of {\it nearby} active galaxies with jet/gas interactions. 
We discuss
the implications on the alignment effect of high redshift  ($\geq$0.7) radio
galaxies. We  conclude that 
although shocks are likely to destroy dust grains
in regions of interaction, dust  might survive in enough quantities 
 to scatter light from the
active nucleus and produce alignment between scattered light and the radio structures.

We propose an observational test to investigate the existence of dust
in shocked regions based on the sensitivity of calcium to  depletion onto dust grains.

\end{abstract}


\section{Introduction}

	The discovery of  high levels of polarization in the extended  continuum of
many  powerful radio galaxies at redshift $z>$1 (Cimatti et al. 1997; Vernet et al. 2001) proves that the alignment between the 
optical  and the radio structures (McCarthy et al. 1987; Chambers et al. 1987) observed in most powerful radio galaxies
has an important contribution of light from a hidden
quasar scattered towards the observer by dust and/or electrons. Several reasons support
that dust is  the scattering agent, rather than electrons (see Cimatti et al. (1998) and
De Young (1998) for a discussion).
This interpretation faces some problems: there is now increasing evidence for the existence of fast 
shocks in the extended
gas of distant radio galaxies  generated during 
the interaction between the radio jet and the ambient gas (e.g. Best, R\"ottgering, Lehnert 2000).
De Young (1998) raised the question of the unlikely survival of dust under these extreme conditions.
The author calculated the survival of dust grains for a wide range of parameters appropriate for the
conditions expected in distant radiogalaxies. De Young found that for most of the cloud configurations the
grains are destroyed as a scattering population by sputtering processes in a time much less that minimum
radio source lifetime of $\sim$10 million yr.

	We investigate in this paper the 
existence of dust in the extended gas of {\it nearby} active galaxies where  
jet/cloud in\-teractions are occurring. There are several low redshift active galaxies  
where evidence for shocks (originated in jet/gas interactions) and dust  in the extended
gas has been found. Cygnus A is an example:   dust
illuminated by the active nucleus  is observed in optical linear polarization maps  which show
a giant bipolar reflection nebula (Tadhunter, Scarrot \& Rolph 1990).  Tadhunter (1991) suggested
that the  components of high velocity gas  discovered  
in the extended narrow line region could be the result of the interaction between the
radio jet and the gas. Another example is 
PKS1932-46. Villar-Mart\'\i n et al. (1998) proposed that the extended gas is
interacting with the radio structures. 
The existence of dust in the extended gas is suggested  by the lack of [CaII]$\lambda\lambda$7291,7324 emission
(see Fig.~4 in Villar-Mart\'\i n et al. (1998)).
The absence of these lines is likely to be due to 
the depletion of calcium  onto dust grains (\S 5).

Therefore, jet/cloud interactions and dust exist in the extended gas of some radio galaxies. However,
in these specific cases it is not clear whether dust {\it is mixed with the shocked gas}.
Grain destruction might have occurred in the shocked gas, but grains could still exist in the non perturbed 
regions.

	We discuss in this paper 
 the case of two active ga\-laxies where the apparent coincidence between
the radio and dust rich structures strongly suggests that the shocked gas is mixed
with dust:  the Seyfert 2 galaxy
NGC1068 (Seyfert 1943) (\S2) and the radio galaxy PKS2152-69 (Westerlund \& Smith 1966) (\S3). 
By applying models appropriate for the shock conditions in NGC1068, we investigate whether dust
grains are likely to survive the passage of the radio jet.   In \S4 we propose an
observational method to search for dust in the shocked (or non shocked) gas of low redshift
active galaxies. Discussion and conclusions are presented in \S5.

\section{NGC1068}

\subsection{Synchrotron emission from the radio jet?}

  The Seyfert 2 galaxy NGC1068  (Seyfert 1943) has been widely
studied at all wavelengths (see for instance Astrophysics and Space Science, v. 248). 
Mid Infrared (MIR) ima\-ges reveal 
a remarkable similarity between  the MIR and the radio structures, 
suggesting that they are spatially
associa\-ted  (Alloin et al. 2000, AL00 thereafter). If this is the case and if the MIR emission is 
produced by dust, then 
  NGC1068 could be an example of an object where  dust has
survived the passage of the radio jet.

The authors identified 
four  structures (in addition to the core) in NGC1068\footnote{Throughout this paper
we will adopt a distance to NGC1068 of 14.4 Mpc  where 1'' corresponds to $\sim$72 pc} at 11.2 and 20.5
$\mu$m, labelled as $a$, $b$, $c$ and $d$ (we adopt the same nomenclature). Regions $a$  (1.8"$\times$4.0") and $b$ 
(1.3"$\times$1.3") are located 
along PA=35 deg to the
northeast at mean
distances from the core of 3.7" and 1.8"
respectively. Regions $c$ (1.3"$\times$1.6") and $d$  (1.3"$\times$1.3") are 
along PA=210  deg to the
southwest at mean
distances from the core of 2.1 and 3.5 " respectively.

	First we have investigated  the possibility that the MIR emission is due to
synchrotron radiation from the radio jet, rather than thermal emission from 
dust.   This is suggested by the close correspondence between the
radio and MIR structures. Jets are rarely observed at MIR   and optical wavelengths 
but some cases
exist, such as the radio galaxy M87 (e.g. Stocke, Rieke \& Lebofsky 1981). For NGC1068 we have 
calculated the expected synchrotron emission at 20.5 $\mu$m  (the emission will be fainter at
11.2 $\mu$m)
from the radio jet   in  two of the bright MIR structures as identified by AL00  ($b$ and
$c$). 
We measured the surface brightness at 4.9 GHz in $b$ and $c$ 
using the radio map of Wilson and Ulvestad (1987). Assuming $S\propto \nu^{-\alpha}$
and using the spectral index   between 4.9 and 15 GHz
measured by
 these authors  in both regions $b$ ($\alpha$=2.1) and $c$ ($\alpha$=1.2), we calculated 
the expected synchrotron emission at 20.5 $\mu$m. These values are upper li\-mits, 
since the spectral index of the synchrotron spectrum is often observed to steepen with
frequency (e.g., Cyg A. Kellermann, Pauliny-Toth \& Williams 1969). The results
(presented in Table 1)   show that synchrotron emission does not make 
a significant  contribution to the MIR emission. We therefore conclude that the MIR extended 
emission in NGC1068 is due to dust (as claimed by e.g. AL00).

\vspace{1cm}
\begin{table*}
\centering
\large
\begin{tabular}{lllllll}
\hline
Aperture & $D$   &  Ext  &   SB$_{4.9GHz}$ &  $\alpha$ & SB$_{20.5}^{synch}$ & SB$_{20.5}^{meas}$ \\ 
	&   ''  & ''$\times$''	  &   Jy arcsec$^{-2}$ &  2.1	 &   Jy arcsec$^{-2}$	&  Jy arcsec$^{-2}$		\\ \hline
b (N-E)	&  1.8  &  1.3$\times$1.3  & 0.031 & 2.1  & 4.3e-7 & 0.65 \\
c (S-W)	&  2.1  &   1.3$\times$1.6 & 0.019 &  1.2 &  6.8e-8 &  0.14\\
 \hline
\end{tabular} 
\caption{Comparison between the predicted surface brightness at 20.5 $\mu$m due
to synchrotron emission (SB$_{20.5}^{synch}$) and the observed surface brightness (SB$_{20.5}^{meas}$)
at 20.5 $\mu$m (ALL00). $D$ is the distance to the core in arcsec. $Ext$ is the projected size.
SB$_{4.9GHz}$ is the surface brightness in
$b$ and $c$ at 4.9 GHz.  $\alpha$ is the spectral index measured by Wilson and Ulvestad (1987) between
 4.9 and 15 GHz. 
Synchrotron emission makes a negligible contribution to the MIR in NGC1068.}
\end{table*}

\subsection{The survival of dust grains}

	Have the dust grains survived the passage of the radio jet in NGC1068 ? 
Is a mechanism of replenishment
required to explain the existence of dust mixed with the shocked gas? 
De Young (1998) studied the response of  dust grains to the passage of the strong shocks associated
with the radio source in an environment where the dust is located in gas clouds of
high density contrast and small
filling factors embedded in   a more diffuse medium. He explored a wide range of
parame\-ters. The factors that influence the evolution of the grains are the initial cloud
gas
density, the cloud size, the initial grain size, and the speed of the shock associated with the radio jet.
Indirect factors arise from the density of the intercloud medium and the presence or
absence of magnetic fields in the clouds. 
We have run similar models with parameters appropriate for NGC1068   and
we have investigated whether dust can survive the passage of the radio jet 
in this galaxy. 

\subsubsection{Model input parameters}

The parameters for NGC1068 are the following:

\begin{itemize}

\item Initial cloud gas density ($n_w$): 
Spectroscopy of the  regions not interacting with the radio jet implies densities of 100 cm$^{-3}$ [Axon et al. 1998]. 
Since this
gas is not affected by the shocks, we have considered this value as the initial
cloud density, before the passage of the radio jet 
(the measured density of the clouds in the region where the jet is
interacting with the gas is 10$^{4.5}$ [Axon et al. 1998]; in our interpretation, such 
high densities
 arise naturally from shock compression)

\item Number density contrast $n_h/n_w$: (where $n_h$  is the  
density of the intercloud medium).   
The measured temperature in the ionised gas is
$T_e\sim$1.15$\times$10$^4$ K (Axon et al. 1998). We expect 
the hot phase temperature to range  from
 10$^6$ to 10$^7$ K  (e.g. Forman, Jones \& Tucker 1985, Shirey et al. 2000,
Netzer \& Turner, 1997). If the two phases 
are in pressure equilibrium, this implies that the number density contrast $n_h/n_w$
(which
is the only intercloud medium parameter that matters
for grain evolution) is in the range 10$^{-2}$ to 10$^{-3}$.

\item Speed of the shock associated with the radio jet $v_s$: 
Axon et al. (1998) report 
 a velocity shift between the split 
components in the optical emission lines of 1500 km s$^{-1}$ at the position
of the interaction between the radio jet and the ionised gas. 
We assume that the velocity of the
clouds accelerated by the shocks is $\sim$750 km s$^{-1}$ (Axon et al. (1998)
deduce shock velocities of at least 700 km s$^{-1}$). In the absence 
of entrainment processes, the cores of the warm clouds will be accelerated
to $v_w  = v_s (\rho_h/\rho_w)^{0.5} \sim v_s (n_h/n_w)^{0.5}$, where $v_w$ is the
final velocity of the warm clouds  and $v_s$ is the shock speed in the hot phase. This implies
shock velocities in the hot phase in the range 7$\times$10$^3$ km s$^{-1}$ and
2$\times$10$^4$ km s$^{-1}$, depending on
the number density contrast.  

\item Clouds size ($R_{cloud}$):
as the uncertainty on this value is high we have considered as 
upper and lower limits for the radius of a cloud as 1000 pc and 1 pc respectively.
We find that the only time the
initial cloud radius makes any difference is in the $B$=0 (no magnetic field) case,
which  is rather unlikely (see below).

\item Dust properties: we  have assumed (as AL00) the MNR  distribution $n(a)=K_a a^{-3.5}$, where $n(a)$ 
is the number density of grains and $a$ is the grain
radius. Upper and lower limits on $a$ were taken as 1000 and 10000 \AA\ respectively .

\end{itemize}

\subsubsection{Models results}

	We show in Table 2 the results of  the models for the range of parameters appropriate for
NGC1068.  The final column indicates the survival time of dust grains of size $R_0$(\AA)
under the conditions considered in that specific model. 
(Grains are considered to be destroyed when the grain radius falls
below one percent of the initial radius, leaving only 10$^{-6}$ of the
initial mass). For the models in the last three rows, the scale for dust destruction
is $>$10$^6$ yr.

Therefore, grains of 1000 \AA\  radius  are 
destroyed in less than 1000 yr
years for the shock speed and density contrast values expected in NGC1068.
Only if there is 
no magnetic field in the cloud, or in the case of 'bare grains' (see below) can small grains
survive more than a million years. 
The large grains (10$^4$ \AA\ ) can survive in low velocity shock
conditions for very long times. Since 
almost all the grains are at the small end of the scale in the MNR distribution,   $\sim$1000 years 
after the passage of the radio jet  most dust should have been destroyed.

\begin{table*}
\centering
\begin{tabular}{llllllll}
\hline
 $n_h/n_w$  &       $v_s$ (km s$^{-1}$) &     $R_{cloud}$ &    Grain $R_0$(\AA)  &    Survival time (yr) \\ 		
\hline   
   10$^{-2}$   &           7x10$^3$     &      1 pc     &       1000        &   9.1$\times$10$^2$     \\ 
10$^{-3}$	&	2x10$^4$	&  "    &   "	&  9.5$\times$10$^2$  \\
  10$^{-3}$     &          2x10$^4$    &      "   &       10000         &     1.6$\times$10$^4$ \\ \hline
10$^{-2}$	&	7x10$^3$	& " &	10000 &   $>$10$^6$	\\
    10$^{-2}$   &         7x10$^3$   &     100 (B=0) &       1000     &        $>$10$^6$ \\ 
    10$^{-2}$    &         7x10$^3$   &    BARE GRAINS   & 1000      &        $>$10$^6$ \\
\hline
\end{tabular}
\caption{Grain survival rates for an internal cloud shock velocity of $v_w$=750 km s$^{-1}$
and the different parameters discussed in the text.}
\end{table*}

	If the outer part of the radio jet interacted with the gas  in region $b$, when
did this interaction happen?  Using  an advance jet speed in the range 0.1$c$-0.01$c$ (e.g.,
Scheuer 1995)\footnote{Based on the
similarities in radio source size and velocity 
between Mrk 3 and NGC1068  Capetti et al. (1999) proposed an upper limit
for the  age of the radio source in 
NGC1068  of $\sim$1.5$\times$10$^5$ yr. Using the projected distance between the outer
edge of the radio lobe and the core ($\sim$6 arcsec) the jet advance speed is $\geq$0.01$c$)}.
and  the distance between the 
outer edge of the jet and region $b$ $\sim$4  arcsec (288 pc in projection),
the interaction should have happened
more than 10$^4$ yr ago. Therefore, according to the model results the dust should have 
been des\-troyed.  However, there is no evidence for dust destruction
in NGC1068.

We propose several explanations:

\begin{itemize}

\item Dust has not been destroyed

\begin{itemize}

\item  The outer radio source did not interact with the gas in regions $b$ and $c$.
On the contrary, the interaction  could have happened very recently (it might still be going on) 
with younger radio structures and there has not been enough time for dust
destruction.

	An upper limit for the time elapsed since the interac\-tion can be calculated
from the cooling time of the shocked gas. If the shocked gas 
is still hot enough
to produce ionizing photons (as  proposed by Axon et al. (1998), see also below), 
the time elapsed since the 
interac\-tion should be shorter than the cooling time.
For shock velocities in the range 500-1000 km s$^{-1}$, the shocked gas will
cool to the recombination temperature of hydrogen in a time-scale $t_{rad,6}$ Myr given by
$t_{rad,6} \sim 1.9~n_w^{-1}~ {v_w}_3^{2.9}$ 
where ${v_w}_3$ is the shock velocity in the clouds in units of 1000 km s$^{-1}$
(Bicknell, Dopita \& O'Dea 1997). For $v_w$=700 km $s^{-1}$ and 
$n_w$=100 cm$^{-3}$, the shocked gas will cool in $\sim$6.7$\times$10$^3$ yr. 
 The
interaction happened, therefore, less than 6.7$\times$10$^3$ yr ago. This might
point to a recent interaction, but it is not possible to assure that it happened
less than 1000 years ago.

\item The magnetic field in the cloud is $B$=0. In this case grains survive the
slower shock speeds, since the cloud cools first and there is no
betatron effect (Spitzer 1941). If  B=0 and one turns on saturated conducti\-vity,
the grains still survive a long time. 
The cloud heats, and 
sputtering increases, but as the cloud heats it also expands,
and the lower number density greatly reduces the sputtering rate.
The small grains are still at 0.73 of their original radius after
2x10$^4$ yr. The case of null magnetic parameter is, however, unlikely.
In all the galaxies we know about, the interstellar
medium is permeated with a magnetic field.    For active galaxies in
particular, where the presence of synchrotron radiation implies a
magnetic field, one would expect a significant magnetic field in clouds near the
nucleus.   We therefore reject this interpretation.
 
\item  The clouds consist of
an external shell of ionised gas and denser cores where the dust has been
protected from the intense UV radiation from the active nucleus.
After the radio jet has hit the
clouds and the shocks have crossed the outer warm ionized layers (with 
$n \sim$100 cm$^{-3}$), the inner dusty cores become exposed.
For the cores, $n_h$/$n_w^{core}$ could be so low that the shocks would  be very
slow and  the grains could survive ($n_w^{core}$ is the gas density in the cores).  
This model has been proposed by Bremer, Fabian \&  Crawford (1997) to explain the alignment 
effect in HzRG as due to scattered light from the hidden quasar, since an intense {\it dust} 
scattered continuum is expected along the radio axis. This is supported by
the remarkable similarity between  the  radio structures and the
J and H band polarized flux  morphologies (due to dust scattered light) in spatial
scales of $\sim$5" from the nucleus  (Packham et al.  1997). 
The enhanced MIR emission at the position of the radio structures also supports
this interpretation (see  \S 2.1.3).

As an example, if the density contrast between the   
hot phase and the dense cores is $n_h$/$n_w^{core}$=10$^{-5}$, a shock propagating at $v_h$=2$\times$10$^4$ km s$^{-1}$
in the hot phase 
would advance with $v_w^{core}$=63 km s$^{-1}$ within the dense cores. The destruction of the
dust grains will be very inefficient with such slow shocks  (De Young 1998).
\footnote{Notice that the emission lines still require shock speeds of $v_w$=750 km s$^{-1}$
and the shocks slow down to $v_w^{core}$ when they enter the molecular core.}
         
For such density contrast, and assuming  $n_h$=0.1 cm$^{-3}$, the density of the 
dense cores is $n_w^{core}$=10$^4$ cm$^{-3}$. 
If this phase is in equilibrium
with the hot phase,  then $T_w^{core}$=10$^2$ K. These properties correspond to a   
molecu\-lar phase.  Lepp et al. (1985) showed that obscuration by  
 gas and dust allows the existence of a stable (cooled by molecules) phase  ($T\sim$10$^{2}$ K)
in the ISM  heated and ionised by a soft X ray continuum.
 A column density of $N_H$$\sim$10$^{22}$ cm$^{-2}$ would be necessary to allow the existence
of this phase. If $n_w$=10$^4$ cm$^{-2}$, this implies that the thickness 
of the  'protective'
shell is $\sim$0.3 pc, which is in good agreement with the minimum assumed total 
cloud sizes (1 pc).

\item In the case of bare grains (i.e. not embedded in
protective clouds, but distributed uniformly trough out a homogeneous hot
circumgalactic medium)  the dust grains
 are very slowly eroded, with the smallest grain (1000 \AA) still being at
0.91 its original radius after 2x10$^4$ yr.  A mechanism to
generate this dust distribution is, however, unknown (De Young 1998).

\end{itemize}

\item Dust has been destroyed and a mechanism of  replenishment is at work 

\begin{itemize}

\item  Dust production
in situ by jet induced
star formation is not viable, since there is no evidence for such a process in
NGC1068. Another
possibility is that dust grains are convected from the dusty
interior of the galaxy outward with the continuously outflowing radio jet. The issue
of grain survival in a hostile environment might, however,  still be a problem 
(De Young 1998).

\end{itemize}

\end{itemize}

\vspace{0.7cm}

{\it In summary, the close association between the extended MIR and radio structures 
in NGC1068 suggest that the shocked gas is mixed with dust. Calculations of dust
survival for NGC1068 suggest that the dust should have been destroyed. There might be a
mechanism that  replenishes the  dust grains, however, there is no evidence for it.
Other possible  explanations for this inconsistency 
are a) the interaction happened recently and there has not been enough time
for dust destruction  b) the clouds contain inner molecular dusty cores where dust destruction
is less likely to happen.}

\subsection{The heating of the dust}

	Why is the dust emission enhanced in the regions crossed by the 
radio jet? 
	  
	One possibility is that  the AGN radiation field is highly anisotropic
(Axon et al. 1998). An alternative explanation is  that, in addition to the continuum
from the active nucleus,  there is another mechanism working along
the radio axis that heats the dust locally to higher temperatures, enhancing
the radiation.

The shocked gas can heat the dust locally via two diffe\-rent mechanisms:
impact with the dust grains and radiative heating.

 The most efficient heating {\it by impact} is due to electrons. In this case,  the temperature
of the dust grains is given (provided that $T_{gas}>>T_{dust}$) by:

$$T_{dust} = [({\frac{3 ~k ~T_{gas}}{m_e}})^{3/2} ~~ {\frac{n ~m_e ~\epsilon}{2~\sigma}}]^{1/4}$$

where $n$ is the density of the shocked gas, $m_e$ is the mass of the electron, 
$\sigma$ is the Steffan-Boltzmann constant, $k$ is the Boltzman constant
and $\epsilon$ a number smaller than 1 than accounts for the fact that
the transfer
of energy from the electrons to the dust grains is not completely efficient
(eq. 4.27, Evans 1994).
The shocked gas can reach tempera\-tures as high as $T_{gas}$=10$^6$K 
(Dopita \& Sutherland 1996). Assuming $n$=10$^{4.5}$ cm$^{-2}$ (see \S 2.2.2), the temperature
of the dust would be $\sim$30 K. Since $\epsilon$ must be $<$1, the heating
we have calculated is  an upper limit and therefore, {\it the heating produced by
collision with the electrons is not important.}  

	A more efficient heating source is the UV radiation from the
shocked gas. Evidence for this radiation in NGC1068 has been discussed by several authors.
 Axon et al. (1998) found 
an increase in excitation of the gas  (despite an increase in density) and
 an excess of continuum along the radio jet. The authors suggest  that in addition to
the central AGN, a  local ionization source is present,  
which is likely to be the ionizing radiation from the shocked gas.

We have checked whether this mechanism  can explain
  the extended MIR emission  in NGC1068. In order to do this, we have calculated
the UV luminosity required to  account for the MIR dust emission
 measured in the  filters
used by ALL00 and compared with the luminosity  produced by 
the shocks (taking into account the higher 
absorption coefficient of dust grains at shorter wavelengths
and the strong UV continuum expected  from incident radiation, we have assumed that the heating
is dominated by photons with wavelength in the spectral range 30-3000 \AA\ ,
Ryter 1996).

	Only a fraction
$f$ of the incident radiation is absorbed by the dust. This radiation is
reemitted in the infrared and only a fraction $\epsilon$ is  in the spectral window
of a given filter. Therefore,
the fraction of incident energy transformed into thermal energy in the filter is:

$\eta = \epsilon ~ f$

	Following the procedure by Alonso-Herrero et al. (1998), we obtain
$\eta \leq$0.04 (since $f\leq$1) for the filter with $\lambda_0$=20.5  $\mu$m and 
   FWHM=1.1 $\mu$m, that is, 
4\% or less of the incident radiation is re-emitted as thermal dust e\-mission within the 
20.5 $\mu$m filter. We obtain $\eta \leq$=0.026  for the filter with $\lambda_0$=11.2  
$\mu$m  and FWHM=0.44 $\mu$m. For these calculations we have assumed that the MIR emission in a given filter
is dominated by dust with temperature such that the maximum flux is emitted at
the central wavelength of the filter; i.e. 145 and 260 K respectively (ALL00 et al. 
conclude that the expected temperature is ~150 in region $c$).

The UV luminosities (L$^{UV}_{11}$, L$^{UV}_{20}$) needed to sustain the MIR luminosities  
measured in regions $b$ and $c$ (L$^{MIR}_{11}$, L$^{MIR}_{20}$) are shown in 
Table 3 (L$^{MIR}_{11}$, L$^{MIR}_{20}$ have been calculated with the flux values
given by AL00 and the properties of the two filters).
Do shocks generate luminosities of this order?

\begin{table}
\centering
\large
\begin{tabular}{llllllll}
\hline
Location &     L$^{MIR}_{20}$   &  L$^{MIR}_{11}$  &  L$^{UV}_{20}$  & 	L$^{UV}_{11}$  	 & L$^{UV}_{shock}$  &	\\  \hline
b (N-E)	&    2.3       &  6.6	&  $\geq$57.5   &   $\geq$253.0   & $\leq$106	\\
c (S-W)	&    2.1      &   0.8  	&   $\geq$52.5   &   $\geq$30.7 &  $\leq$130	\\
\hline
\end{tabular}
\caption{Comparison between the UV luminosities required to sustain the observed MIR luminosities and the UV luminosity
generated by the shocked hot gas.
All luminosities are given in units of 10$^{41}$erg s$^{-1}$. None of the luminosities are 
 monochromatic, but the luminosities in  the filters discussed in the text.  L$^{UV}_{20}$ and
L$^{UV}_{11}$ are the required UV luminosities to sustain the observed MIR luminosities L$^{MIR}_{20}$ and L$^{MIR}_{11}$, 
for the conversion efficiency 
calculated in the text. L$^{UV}_{shock}$ is the UV luminosity produced by the shocks ($v_s$=700 km s$^{-1}$). }
\end{table}

 The UV ionizing radiation field (over 4$\pi$ sr) 
produced by a shock of velocity $v_s$ propagating in a medium of (unshocked) density $n$ 
is given by  eq. 4.1 in Dopita \& Sutherland 1996:

$$L^{UV}_{shock}  = 1.11 \times 10^{-3} ~ {[\frac{v_s}{100~km~s^{-1}}}]^{3.04} ~ \frac{n}{cm^{-3}} ~erg ~cm^{-2} ~s^{-1}$$

For $v_s$=700 km s$^{-1}$ and $n=$100 cm$^{-3}$ (see \S 2.2.2)  
the total UV luminosity emitted by the shocked gas in
regions $b$ and $c$  is given in Table 3. To calculate the area occupied by the shocked gas, we  
have assumed that $b$ and $c$ have cylindrical geometry. The height and
diameter of the cylinder were taken as the projected sizes of regions $b$ and $c$ indicated in Table~1.
We obtain $Area_b$= 2.6$\times$10$^{41}$ cm$^{-2}$ and $Area_c$= 3.2$\times$10$^{41}$ cm$^{-2}$ (projected). 
The results of these calculations (Table 3) show that the shocks produce
an intense continuum that is likely to contribute to the dust heating. According to these
results, the shock continuum is enough to explain the MIR dust emission in $c$ and the 
20$\mu$m emission in $b$. However, our calculations
assume that 
the whole area of the cylinder is filled with shocked gas, which
is unlikely  and that all the photons
 are intercepted by the dust. This is certainly not the case, since part of the photons
ionize the gas and part escape unabsorbed (Axon et al. 1998). Therefore,  L$^{UV}_{shock}$ is an upper limit
and  an additional
heating source is  probably present. This is actually needed  to explain the 11$\mu$m flux measured in $b$. 
The active nucleus can provide that extra heating. 
In order to subtain the 11$\mu$m flux in region $b$ 
we need   $L^{UV} \geq$7$\times$10$^{44}$ erg s$^{-1}$ from the active nucleus.
Using the information on the observed  total spectrum of NGC1068 (Pier et al. 1994), Marco \& Alloin (2001) 
 obtained a lower limit of  
$L^{UV} \geq$4$\times$10$^{44}$ erg s$^{-1}$. Therefore, the AGN is likely to provide 
the additional heating source.

 We conclude that 
{\it  the shocked gas generates an intense UV continuum that 
enhances the MIR emission in the regions overlapping with the radio
structures. The AGN continuum also
 contributes to the dust heating.}

Another possibility   mentioned in the previous section is that the radio jet
has exposed the inner dusty cores of the clouds.  In this case, the UV photons must be
generated elsewhere, in the much faster moving emission line gas. 
The calculations done above are also valid
in this case and both the shock and the AGN continuum are likely to contribute to the heating
of the dust in the cores of the clouds.

\section{PKS2152-69 ({\small z} = 0.028)}

 The interaction between the radio jet and 
a high ionization cloud
at a projected distance of 8 kpc from the nucleus has been widely studied
(e.g. Tadhunter et al. 1987; Fosbury et al. 1998). The cloud lies approximately 
along the radio axis  and is believed to
be hit by the radio jet.   The existence of dust in the cloud is confirmed
by the detection of  a very blue optical to near-UV continuum with 10 per cent
polarization and the E-vector perpendicular to the position angle of the nucleus-high
ionization cloud axis (di Serego Alighieri et al. 1988).
The authors explained the continuum and spectropolarimetric properties of the
 cloud as the
result of scattering by dust in the cloud of beamed radiation from the nucleus. 

	Therefore, this appears as a clear example where shocks (produced by jet/cloud
interactions) do not prevent 
dust from surviving in sufficiently large quantities, so it can still
 scatter the nuclear light 
and cause the alignment between the optical and radio structures. 

	This does not imply, however, that shocks have not des\-troyed the dust.
High resolution radio maps and HST optical images (Fosbury et al. 1998) have shown that  {\it part} of
the high ionization cloud overlaps with a bright radio component. 
 The optical continuum emitted by the  overlapping region 
is redder and fainter than the continuum emitted by the 
rest of the cloud. This suggests that  the dust
grains res\-ponsible for the scattering have different properties.
The change of continuum colours could be explained if the
small dust grains (easier  to  destroy) have been 
destroyed by the shocks in the region of the cloud that is interacting
with the radio structure. The scattering would be dominated
by large grains, which produce a redder reflected 
continuum. Small grains dominate the scattering
in the rest of the cloud (as proposed by Magris, Binette \& Martin 1992)
and therefore the scattered continuum is bluer.
Spatial spectropolari\-metric  information  across the cloud
and/or the study of the continuum colour variation   would
confirm whether this interpretation is valid or another explanation is required.


	{\it In summary,  PKS2152-69 is a radio galaxy that presents close
alignment between the optical and radio structures. The aligned continuum
is due to radiation from the active nucleus scattered by dust,
  in spite of the strong interaction between the radio structures and the cloud.
In this case, the observations suggest that at least the small grains have  been
destroyed as a consequence of the interaction  between the gaseous cloud
and the radio jet.  The dominant source of aligned
continuum is the region of the cloud that apparently has not been 
shocked.}

\subsection{Discussion}

	The case of NGC1068 implies that dust can survive under strong shock conditions,
contrary to what we expect from our modeling.  Possible explanations require
recent interactions or  dense molecular dusty cores exposed to the AGN radiation field. 
The close correspondence between the radio and dusty structures in this galaxy
supports the model by Bremer, Fabian \& Crawford (1997)  for the alignment effect in 
high redshift radio galaxies
as a consequence of scattered light. In this model  the radio jet
has exposed the inner dusty cores of the clouds and the contribution of scattered light
is enhanced.  An alternative possibility is that shocks
heat the dust radiatively and enhance the MIR emission along the jet.

The case of PKS2152-69  suggests that shocks can indeed destroy small dust
grains  (although spatially resolved spectropolarimetric observations of the high
ionization cloud are needed to confirm this).  However, dust survives in enough quantities in 
adjacent (or distant) regions 
to produce the appa\-rent alignment between the radio and optical 
structures.

	Both cases show that the interaction between the radio jet and the ambient gas
has not prevented dust from surviving in enough quantities to scatter light from the
active nucleus and produce alignment between scattered light and the radio structures.
This could also be the case in high redshift radio galaxies.

\section{Searching for dust in shocked regions}

	Another way to investigate the survival of  dust in shocked regions 
is by studying the
line emission from species sensitive to depletion, such
 as calcium and iron. In the ISM 99\% of calcium and iron are locked 
in the dust grains, rather than in the gaseous phase.

Villar-Mart\'\i n et al. (1996, 1997) (based on the idea proposed by Ferland 1993) showed that  in the absence of dust, the 
 doublet lines [CaII]$\lambda$7291,7324  should be easily detectable
(stronger than [OI]$\lambda$6300) in   {\it photoionised} nebula 
around nearby radio galaxies
(such that [OIII]$\lambda$5007/H$\beta \leq$10). The weakness (or absence)
of these lines implies the existence of dust mixed with the ionised gas if
the dominant ionization mechanism is {\it AGN photoioni\-zation}, whether shocks are
present or not. This diagnostic method is not reliable when applied to AGN photoionised
nebulae at higher redshift ($z\geq$1). There is evidence (Gondhalekar, Gondhalekar \& Wilson 1988; 
Francis 1993) 
that the AGN continuum is harder 
at higher redshifts. With the expected power law index  
($\alpha$=-1,  Villar-Mart\'\i n, Tadhunter \& Clark 1997
instead of $\alpha$=-1.5 as assumed for low redshift radio galaxies) the [CaII] emission 
lines should be very weak, unless for low ionization parameters. The non detection of [CaII]
emission at high redshift is therefore no so straight forward to interpret in terms
of depletion by dust.

Over the last few years, the dominant emission line mechanism (AGN photoionization
vs. shock related processes)
in high redshift radio galaxies  has been a source of debate   (e.g Villar-Mart\'\i n
et al. 1997,  Best, R\"ottgering \& Lehnert 2000; de Breuck et al. 2000). Both mechanisms are likely to play
a role.

	We investigate here the applicability of the calcium test as a diagnostics for  the existence of dust
under shock ionization conditions.
We have run a set of shock and shock+precursor models with the MAPPINGS III code, which is an enhanced
version of  the MAPPINGS II code (Dopita \& Sutherland 1996). MAPPINGS III is designed to handle higher
temperatures (up to 10$^8$ K) and hence faster shocks. The models considered here range from 250 to 
800 km s$^{-1}$
which are appropriate for the physical conditions
of gaseous nebula around radio galaxies
(the maximum ve\-locity considered by Dopita \& Sutherland was 500 km s$^{-1}$). 
We have predicted the
strength of the [CaII]$\lambda$7291 emission line relative to other lines 
in the absence of dust. The abundances are solar and non depleted by dust. In Fi\-gure 1 we plot the variation of 
the [CaII]$\lambda$7291 flux (the strongest
line in the doublet) 
relative to other emission lines with
 shock velocities (250-800 km s$^{-1}$). The two sequences (solid and dashed lines)
correspond to magnetic parameters $\frac{B}{n^{1/2}}$=1 and 4 respectively. The diagram on the left shows
the pure shock models and the diagram on the right shows the shock+precursor models.
The models show that over this velocity range [CaII]$\lambda$7291 should be stronger 
that [OIII]$\lambda$4363 and  HeII$\lambda$4686 and, for velocities below
500 km s$^{-1}$, the line should be stronger or comparable to [OI]$\lambda$6300.

\begin{figure*}
\includegraphics{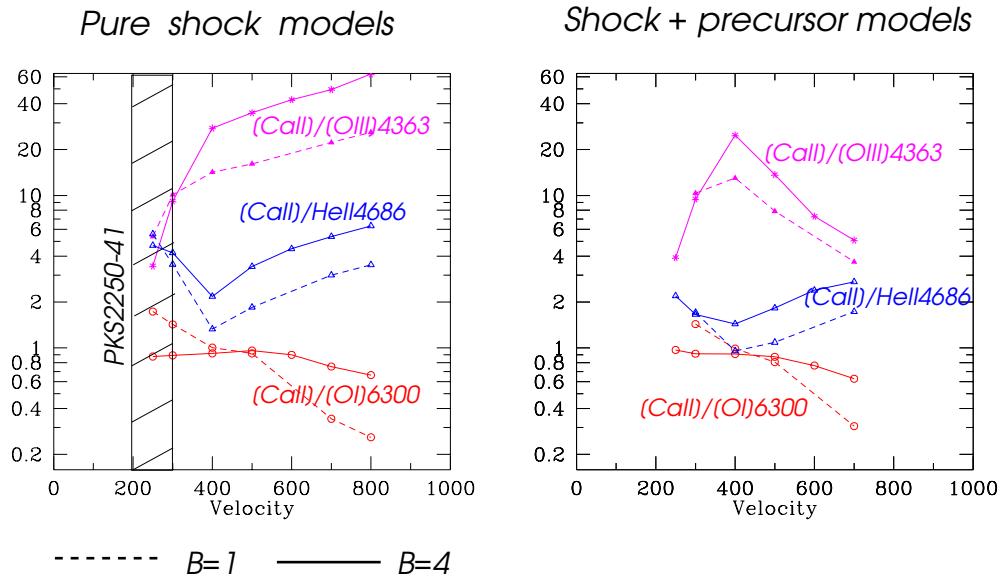}
\vspace{3in}
\caption{Variation with shock velocity (in km s$^{-1}$ of the  flux of [CaII]$\lambda$7291 {\it relative} to HeII$\lambda$4686, 
[OIII]$\lambda$4363 and [OI]$\lambda$6300
for shock models with no (left) 
and  with (right)  precursor contribution. Each line in the diagram represents the ratio Magnetic parameter $\frac{B}{n^{1/2}}$ is 
1 (dashed lines) and  4 (solid lines) $mu$G cm$^{-3/2}$. 
 The shadowed region shows the velocity
range of the models that reproduced the optical line ratios of PKS2250-41. For all models [CaII]$\lambda$7291 should 
be stronger than [OIII]$\lambda$4363 and HeII$\lambda$468. The absence of weakness of the line would imply that the shocked gas is
mixed with dust and therefore, the shocks have not destroyed the dust.}
\end{figure*}

{\it I.e., the [CaII]$\lambda$7291
 line should be detectable in the extended gas of radiogalaxies (and the NLR of Seyferts) with shocks
(whatever the ionization mechanism is), stronger than  HeII$\lambda$4686 and [OIII]$\lambda$4363
(and similar to or stronger than [OI]$\lambda$6300). The absence  or weakness of the line 
 would imply that dust is mixed with the shocked
gas and the shocks have not destroyed
the dust.  The  applicability of this test to high redshift radio galaxies
where AGN photoionization is the dominant ionization mechanism is not obvious, except
for low values  of the ionization parameter.} 

	This test was applied to the high ionization cloud in  PKS2152-69 (Villar-Mart\'\i n et al. 1997).
The lack of [CaII] emission
 was interpreted by the authors 
as evidence for the existence of dust mixed
with the ionized gas. This adds evidence to the survival of dust in the cloud.

	An interesting case to compare the models with is 
 the radio galaxy PKS2250-41 ($z=$0.31), where it has been possible to isolate
the shocked and the non shocked gas. Clark et al. (1997) found clear evidence 
for a strong interaction (and shocks) between
the radio jet and the extended gas of this radio galaxy. 
Villar-Mart\'\i n et al. (1999) made the spectral decomposition of the emission lines 
emitted by the interac\-tion region. This led to the kinematic resolution of the shocked 
gas (FWHM~600-900 km/s) and  the non perturbed gas (FWHM$\sim$150 km/s) in the main  
emission lines. The line ratios measured for the shocked gas were consistent
with shock ionization models with velocities $\sim$200-300 km s$^{-1}$.
As Fig.~1 shows, if the shocked gas is dust free,  [CaII]$\lambda$7291
should be stronger than HeII$\lambda$4686, [OIII]$\lambda$4363 and similar
or stronger than [OI]$\lambda$6300. Since these two lines have been detected
in the shocked gas of PKS2250-41 (Villar-Mart\'\i n et al. 1999),  broad [CaII]$\lambda$7291
(FWHM$\sim$600-900 km $^{-1}$)   should also be detected. 
If  this is the case, we shall conclude that the gas is dust free (calcium is
not depleted) and 
shocks have destroyed the dust.
 The non detection of the line 
will imply that the {\it shocked gas} contains dust.

\section{Summary and conclusions}

The goal of this paper  has been to study the survival of dust in low redshift active galaxies 
where shocks have been induced by the passage of the radio jet and link the conclusions with the
alignment effect in high redshift radio ga\-laxies.

The close correspondence between the extended MIR and radio structures in NGC1068 suggests that dust exists
in the regions shocked by the radio jet. We show that, in addition to the AGN  radiation, the
continuum generated by the shocks contributes to the dust heating in these regions, enhancing in this way
the
MIR emission. 

Models appropriate for NGC1068 show that small dust 
grains should have been destroyed in less than 1000 years. We propose several explanations for the dust
survival, such as a recent interaction. Another possibility is that
the  clouds consist of
an external shell of ionised gas and dense mole\-cular   dusty cores. The radio jet
has exposed the  cores to the intense UV radiation field of the nucleus and the MIR emission is enhanced
along the radio axis. This  model was proposed by Bremer, Fabian \& Crawford (1997)  
to explain the enhancement of scattered light along the radio axis in 
high redshift radio galaxies.

The close radio/optical alignment in the radio galaxy PKS2152-69 (due to dust scattered light)
and the evidence for a strong interaction between the radio jet and the gas,
prove that jet induced shocks and radio/optical alignment due to dust scattered
light are compatible. Based on the different continuum colour observed between the
shocked and non shocked regions, we suggest that small dust grains have been destroyed in the region
of the interaction. This  has not prevented the alignment between the dust scattered
continuum and the radio structures to be observed.

In summary,   there can be high redshift radio galaxies where the cloud properties are such that
dust destruction is prevented (for instance  dense dusty cores) or where the interaction 
happened very recently and there  has not been enough time
for the dust to be destroyed.
Another possibility is that, as in PKS2152-69,
dust has been destroyed in the shocked gas, but  regions adjacent (or distant) to the radio structures
contain enough dust to account for the observed alignment. 

Our conclusion is that jet/cloud interactions and the alignment effect due to dust scattered light are
compatible.

We propose to use the [CaII]$\lambda$7291 line to investigate the existence of dust mixed
with the shocked gas in  active galaxies  (radio galaxies, Seyferts) where shock ionization (with or without precursor contribution)
  or AGN photoioni\-zation 
dominate the emission line processes.
In the absence of dust, [CaII]$\lambda$7291 from the shocked gas 
should  be stronger than [OIII]$\lambda$4363  and HeII$\lambda$4686
(both lines often detected). 
 The absence of the [CaII] line would suggest that the calcium is depleted
onto dust grains in the shocked gas. 
PKS2250-41 appears as a  good candidate to apply this test, because the shocked and the non perturbed gas
have already been resolved kinematically. 
If shocks have destroyed the dust, broad (FWHM$\sim$600-900 km s$^{-1}$) [CaII]$\lambda$7291 should be
detected.

\section{Acknowledgments}
We thank an anonymous referee for useful comments that helped to improve the paper.
We thank A. Wilson for providing the radio map of NGC1068. MVM thanks 
Carlos de Breuck for useful discussions. The work of LB was supported by the CONACyT grant 32139-E.

\end{document}